\documentclass[aps,prl,twocolumn,floatfix]{revtex4}
\usepackage[latin1]{inputenc}
\usepackage{amsfonts}
\usepackage{amssymb}
\usepackage{amsmath}
\usepackage{calc}
\usepackage{graphicx}
\usepackage{epsfig}
\usepackage{bm}
\usepackage{ulem}
\usepackage{setspace}
\usepackage{subfigure}

\begin{document}
%%%%%%%%%%%%%%%%%%%%%%%%%%%%%%%%%%%%%%%%%%%%%%%%
%%%%%%%%%%%%%%%%%%%%%%%%%%%%%%%%%%%%%%%%%%%%%%%%

\title{Fock-state view of weak-value measurements and implementation with photons and atomic ensembles}
\date{\today}
%\pacs{}%03.67.Hk, 03.67.Mn, 42.50.Md, 76.30.Kg}
\author{Christoph Simon$^1$ and Eugene S. Polzik$^2$}
\affiliation{ $^1$Institute for Quantum Information Science
and Department of Physics and Astronomy, University of
Calgary, Calgary T2N 1N4, Alberta, Canada\\$^2$ Niels Bohr
Institute, Danish Quantum Optics Center - QUANTOP,
Copenhagen University, Blegdamsvej 17, 2100 Copenhagen
{\O}, Denmark}

\begin{abstract}
Weak measurements in combination with post-selection can
give rise to a striking amplification effect (related to a
large ``weak value''). We show that this effect can be
understood by viewing the initial state of the pointer as
the ground state of a fictional harmonic oscillator,
helping us to clarify the transition from the weak-value
regime to conventional dark-port interferometry. We then
describe how to implement fully quantum weak-value
measurements combining photons and atomic ensembles.
\end{abstract}

\maketitle

In 1988 Aharonov, Albert and Vaidman discovered that weak
measurements in combination with pre- and post-selection
give rise to a striking amplification effect, e.g. giving a
``weak value'' greater than 100 for the spin of a spin-1/2
particle \cite{AAV}. Such weak-value measurements have been
studied in various contexts, including the foundations of
quantum mechanics \cite{foundations}, superluminal light
propagation in dispersive materials \cite{superluminal},
polarization effects in optical networks \cite{networks},
and cavity QED \cite{wiseman}. Most recently they were used
for precision measurements \cite{precision}, spurring
further research in that direction \cite{brunner,starling}.

The standard scenario of weak-value measurements involves a
system and a pointer that interact via a Hamiltonian
$H=\chi \sigma_z P$, where $\sigma_z$ is a spin-like
observable for the system and $P$ is a momentum-like
observable for the pointer. If the interaction is weak, and
if the initial and post-selected states of the system are
almost, but not quite, orthogonal, the final state of the
pointer is displaced in $X$ (the canonical variable
conjugate to $P$) by an amount that is much greater than
one would naively expect based on the interaction strength.
In the first part of the present paper we describe an
instructive way of understanding this effect. The initial
state of the pointer can be seen as the ground state of a
fictional harmonic oscillator, while the state of the
pointer that is prepared by the post-selection is a
superposition of the ground and first excited states, with
coefficients that depend on the interaction strength and on
the post-selection measurement. This gives a simple view of
the relationship between the weak-value regime and other
measurement techniques.

This new perspective on weak-value measurements is very
helpful when thinking about possible implementations of
such measurements combining photons and atomic ensembles,
which is the second topic of the present paper. So far weak
values were primarily observed with light, where typically
polarization plays the role of the system and the external
degrees of freedom of the photons play the role of the
pointer
\cite{foundations,superluminal,networks,precision,morelight}.
These optical experiments can be understood using the weak
value formalism, but also in purely classical terms.
Recently there have been proposals to implement weak values
with quantum dots \cite{qdots}. Here we propose to
implement weak values in a fully quantum way via the
coupling of light to atomic ensembles. The interaction of
light with atomic ensembles has been widely used in quantum
information processing \cite{hammerer}, implementing
important paradigms such as quantum memories \cite{memory}
and quantum teleportation \cite{teleport}, making atomic
ensembles very attractive for the realization of quantum
repeaters \cite{briegel,dlcz,sangouard,dlcz-exp}. We show
that weak-value experiments can be implemented by a
modification of the spontaneous Raman scattering experiment
that lies at the heart of the well-known
Duan-Lukin-Cirac-Zoller (DLCZ) quantum repeater protocol
\cite{dlcz,dlcz-exp}.

{\it Fock-state view of weak-value measurements.} As is
customary in the weak-value literature, let us consider the
case where the initial wave function of the pointer is a
Gaussian, $\psi_0(x)=e^{-\frac{x^2}{2 w^2}}w^{-\frac{1}{2}}
\pi^{-\frac{1}{4}}$, where $x$ is the variable
corresponding to the operator $X$ and $w$ is the width of
the Gaussian. Any Gaussian can be seen as the ground state
of a fictional harmonic oscillator Hamiltonian. That is,
one can define an annihilation operator
$a=\frac{1}{\sqrt{2}}(\frac{X}{w}+iw P)$ (satisfying
$[a,a^{\dagger}]=\openone$) which annihilates the initial
state, $a|\psi_0(x)\rangle=0$. One can then identify the
initial pointer state with the ground state of the thought
harmonic oscillator, $|\psi_0(x)\rangle \equiv |0\rangle$.
The Hamiltonian $H=\chi \sigma_z P$ can be rewritten in
terms of $a$ and $a^{\dagger}$ as $H=-i
\frac{\chi}{\sqrt{2}w} \sigma_z (a-a^{\dagger})$.

Suppose that the system is initially prepared in the state
$|x+\rangle$, which satisfies $\sigma_x |x+\rangle
=|x+\rangle$ and $\sigma_z|x+\rangle=|x-\rangle$, where
$|x-\rangle$ is the other eigenstate of $\sigma_x$,
satisfying $\sigma_x |x-\rangle=-|x-\rangle$. Then the time
evolution of the system and the pointer for a weak
interaction is given by
\begin{eqnarray}
e^{-iHt}|x+\rangle|0\rangle \approx (\openone -
iHt)|x+\rangle|0\rangle=|x+\rangle|0\rangle+\kappa
|x-\rangle |1\rangle, \label{entstate}
\end{eqnarray}
where $\kappa=\frac{\chi t}{\sqrt{2} w}$ is supposed to be
much smaller than 1, and $|1\rangle \equiv
a^{\dagger}|0\rangle$ is the first excited state of the
fictional harmonic oscillator. The interaction has created
an entangled state between the system and the pointer. If
$\kappa \ll 1$ the entanglement is very weak, and the
effect of the interaction may be hard to detect. In
particular, the probability to detect the pointer in the
state $|1\rangle$ is $\kappa^2$.

The weak-value protocol proceeds by measuring the state of
the system in a basis that is close, but not identical, to
the $\sigma_x$ basis, and post-selecting those results for
which the measured state is almost orthogonal to the
initial state $|x+\rangle$. That is, we project the system
onto a state that can approximately be written as $\phi
|x+\rangle+|x-\rangle$, where $\phi \ll 1$ is the small
angle between the preparation and post-selection bases. The
resulting state of the pointer is seen from Eq.
(\ref{entstate}) to be
\begin{equation}
\phi |0\rangle + \kappa |1\rangle, \label{pointerstate}
\end{equation}
showing that the probability for the projection to occur is
$\phi^2+\kappa^2$. The usual weak-value regime corresponds
to the conditions $\kappa \ll \phi \ll 1$. In this case the
probability for a successful post-selection is
approximately equal to $\phi^2$, and the pointer state can
be written in renormalized form as $|0\rangle +
\frac{\kappa}{\phi}|1\rangle$.

By applying the creation operator $a^{\dagger}$ to the
ground state wave function one finds the wave function of
the state $|1\rangle$,
$\psi_1(x)=\frac{\sqrt{2}x}{w}\psi_0(x)=-\sqrt{2}w \frac{d
\psi_0(x)}{dx}$. As a consequence, the wave function of the
final pointer state is
\begin{equation}
\psi_0(x)-\frac{\kappa}{\phi} \sqrt{2}w
\frac{d\psi_0(x)}{dx} \approx
\psi_0(x-\sqrt{2}w\kappa/\phi),
\end{equation}
which clearly shows that the wave function of the pointer
is displaced by an amount that is inversely proportional to
the small parameter $\phi$. This corresponds to a large
``weak value'' of the system observable $\sigma_z$ of
$\frac{1}{\phi}$. This terminology is due to the fact that,
for the Hamiltonian $H=\chi \sigma_z P$ and in the absence
of post-selection, the displacement of the pointer would be
determined only by $\kappa$, $w$ and the initial value of
$\sigma_z$. We see now that post-selection allows for much
larger displacements. This approach can be used to amplify
the effect of the small interaction ($\kappa$) at the
expense of a small post-selection probability ($\phi^2$).

The amplified displacement is obtained in the regime where
$\frac{\kappa}{\phi} \ll 1$, so that the final wave
function of the pointer is still approximately a Gaussian.
However, Eq. (\ref{pointerstate}) also allows us to easily
understand other regimes in which the usual weak-value
description no longer applies. In particular, as $\phi$ is
decreased and becomes smaller than $\kappa$, the final
pointer wave function is increasingly dominated by
$\psi_1(x)$, which has the typical ``derivative'' shape.
This regime was recently discussed in the weak-value
context in Ref. \cite{starling}. The limiting case of
$\phi=0$ corresponds to observing the dark port in
conventional interferometry.

It is an interesting question under what exact conditions
the choice of $\phi \gg \kappa$ vs. $\phi \lesssim \kappa$
is advantageous for  measuring the small parameter
$\kappa$. The weak-value regime ($\phi \gg \kappa$) seems
to be advantageous when $\kappa$ is so small that dark-port
detections (which occur with a probability of order
$\kappa^2$) are dominated by background noise, i.e.
$\kappa^2 \ll \beta$, where $\beta$ is the noise level,
whereas $\phi^2$ can be made greater than $\beta$. Note
that the weak-value regime still corresponds to $\phi \ll
1$. Increasing $\phi$ to of order one corresponds to the
conventional ``bright-port'' regime, where there is no
post-selection, but also no amplification effect.
Weak-value measurements have the potential to outperform
``bright-port'' interferometry in certain cases, for
example for the measurement of longitudinal phase shifts in
the presence of alignment errors, see Ref. \cite{brunner}.
Weak-value techniques thus correspond to an interesting and
potentially useful intermediate regime between bright-port
and dark-port interferometry.

\begin{figure}
\epsfig{file=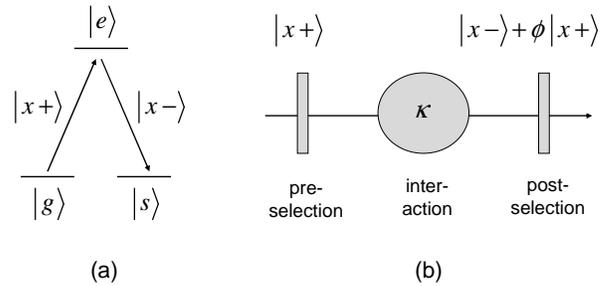,angle=270,width=0.9\linewidth}
\caption{Implementation of weak-value measurements with
photons and atomic ensembles. (a) An incoming photon has a
small probability amplitude $\kappa$ to be scattered from
the polarization state $|x+\rangle$ to the state
$|x-\rangle$, causing the atomic ensemble to go from the
state $|0\rangle$ of Eq. (4) to the state $|1\rangle$ of
Eq. (5). This creates an entangled state between the photon
and the ensemble, see Eq. (1). (b) Photons are observed in
the forward direction and projected onto the almost
orthogonal polarization state $|x-\rangle + \phi
|x+\rangle$, where $1 \gg \phi \gg \kappa$. This projects
the atomic ensemble onto the state $\phi |0\rangle + \kappa
|1\rangle$, which is the initial atomic state displaced by
an amount that is proportional to $\frac{\kappa}{\phi}$,
see Eq. (3). This amplified displacement of the atomic
state can be detected by ``atomic homodyne detection'', see
text.} \vspace{-0.15cm} \label{separated}
\end{figure}

{\it Implementation with photons and atomic ensembles.} It
is possible to realize an interaction Hamiltonian of the
form $H=\chi \sigma_z P$ through the off-resonant
interaction of light with atomic ensembles
\cite{hammerer,memory,teleport}. However, we have learned
from the above discussion that the key point is in fact the
creation of the entangled state Eq. (\ref{entstate}). This
can be done most directly through spontaneous Raman
scattering, similarly to the DLCZ quantum repeater proposal
\cite{dlcz}, see Figure 1. The ``system'' is realized by
the polarization states of a photon, where $x+$ is the
polarization of the ``write'' photon, whereas the $x-$
polarization corresponds to the Raman-scattered ``Stokes''
photon. The scattering of the photon from $x+$ to $x-$
polarization is accompanied by the creation of an atomic
excitation. The atomic system, which is initially in the
state
\begin{equation}
|0\rangle \equiv |g\rangle_1 |g\rangle_2 ...
|g\rangle_{N_A}, \label{zero}
\end{equation}
with all $N_A$ atoms in the ground state $g$, makes a
transition to the state
\begin{eqnarray}
|1\rangle \equiv \frac{1}{\sqrt{N_A}}\left( |s\rangle_1
|g\rangle_2 ... |g\rangle_{N_A}+...+ |g\rangle_1...
|g\rangle_{N_A-1} |s\rangle_{N_A}\right), \label{one}
\end{eqnarray}
which is the symmetric state with one atom in the state
$s$. It is again possible to see the states $|0\rangle$ and
$|1\rangle$ as the ground and first excited states of a
harmonic oscillator, where the number of excitations in
Fock space now corresponds to the number of atoms in $s$.

This can best be seen in two steps \cite{hammerer}. One
begins by introducing a quasi-spin operator
$\sigma_x=|g\rangle \langle g|-|s\rangle \langle s|$ for
each atom, with corresponding expressions for $\sigma_y$
and $\sigma_z$. The associated collective spin operators
$J_x, J_y, J_z$, where $J_x=\frac{1}{2} \sum_{k=1}^{N_A}
\sigma_x^{(k)}$ etc., fulfill commutation relations
$[J_y,J_z]=iJ_x$. The state $|0\rangle$ of Eq. (\ref{zero})
is the eigenstate of $J_x$ with maximum eigenvalue $N_A/2$.
For the phase space region in the vicinity of this state
(i.e. as long as there are just a few atoms in $s$) one can
then employ the Holstein-Primakoff approximation and
introduce new canonical variables $X=J_y/\sqrt{N_A/2}$ and
$P=J_z/\sqrt{N_A/2}$ that satisfy $[X,P]=i$
\cite{hammerer}. The angular momentum ladder operator
$J_-=J_y-i J_z$, which creates excitations in $s$, is then
proportional to the harmonic-oscillator creation operator
$a^{\dagger}=\frac{1}{\sqrt{2}}(X-iP)$, and the states of
Eqs. (\ref{zero}) and (\ref{one}) are related by $|1\rangle
= a^{\dagger}|0\rangle$ as before.

Starting with one photon in the state $|x+\rangle$ and the
atomic ensemble in the state $|0\rangle$, spontaneous Raman
scattering with an amplitude $\kappa$ will create the
entangled state of Eq. (\ref{entstate}). In typical
experiments \cite{dlcz-exp,sangouard} one detects the
Stokes photon, corresponding to a projection onto the
$|x-\rangle$ state, which furthermore projects the atomic
ensemble onto the state $|1\rangle$. In fact, the idea of
the protocol of Ref. \cite{dlcz} is to detect a Stokes
photon that could have come from either of two ensembles,
thus creating a single, but delocalized, atomic excitation,
which forms the basic unit of entanglement for the quantum
repeater protocol. It is important in this context to
distinguish the Stokes photons from the write photons, i.e.
to perform a very accurate projection onto $|x-\rangle$.

Now suppose that our task is not to implement a quantum
repeater, but to measure $\kappa$, which might be very
small in some situations. One possible approach is to
directly detect the Stokes photons as before. However, this
may be impossible for very small $\kappa$ because of
various sources of background noise. The alternative,
weak-value, approach is to detect light in the forward
direction, but to project not exactly onto the Stokes mode
$|x-\rangle$, but onto a superposition state $\phi
|x+\rangle + |x-\rangle$ which has a slight admixture of
the write beam mode, see Figure 1(b). This will project the
atomic ensemble onto a superposition state of the form
$\phi |0\rangle + \kappa |1\rangle$ as before. Generating
such superposition states for the atomic ensemble could
also be a motivation in itself for this approach. The
weak-value regime corresponds to $1 \gg \phi \gg \kappa$,
where $\phi^2$ is chosen such that it is above the noise
level. In the $x$-representation the atomic state will have
the form of Eq. (3) with $w=1$, i.e. it corresponds to the
vacuum state displaced by $\frac{\sqrt{2}\kappa}{\phi}$.

In typical experiments the write beam contains more than a
single photon. However, this does not change the principle
of the proposed implementation, as long as the probability
of scattering a photon into the polarization mode $x-$ and
thus exciting the atomic state $|1\rangle$ remains small.
In the above discussion one can simply replace the
single-photon state $|x+\rangle$ by the state
$|N\rangle_{x+}|0\rangle_{x-}$, which describes $N$ photons
of polarization $x+$, and the single-photon state
$|x-\rangle$ by the state $|N-1\rangle_{x+}|1\rangle_{x-}$,
in which a single photon has been scattered into the $x-$
polarization. The final detection of a single photon is
described by the annihilation operator $\phi a_{x+} +
a_{x-}$, where $a_{x+}$ describes the annihilation of a
photon of $x+$ polarization etc. The presence of $N$
photons enhances both the scattering amplitude and the
amplitude of detecting a photon from the write beam by a
factor of $\sqrt{N}$, but the final atomic state is the
same as before.

In order to explicitly show the displacement of the atomic
pointer state, one has to measure the distribution of
results for the observable $X=J_y/\sqrt{N_A/\sqrt{2}}$.
This can be done by first applying a $\pi/2$ pulse to the
atomic state, followed by a measurement of the population
difference in the levels $g$ and $s$ (i.e. $J_x$). The
latter measurement can be performed using a quantum
non-demolition (QND) interaction with off-resonant light
followed by homodyning of the transmitted light
\cite{hammerer}. Note that the described procedure can be
seen as a homodyne detection of the atomic state, where the
$\pi/2$ pulse corresponds to the beam splitter in
conventional homodyne detection. The population differences
between the states $|0\rangle$ and $|1\rangle$ of Eqs. (4)
and (5), for example, are too small to be detected directly
through the QND interaction, but the two states are well
distinguishable with the described ``atomic homodyning''
technique. The precision of this measurement technique for
$X$ is sufficient to detect displacements that are smaller
than the width of the vacuum wave packet
\cite{hammerer,appel}.

We have described an instructive new perspective on
weak-value measurements and proposed a feasible, fully
quantum experimental implementation with photons and atomic
ensembles. In the future we intend to explore the
application of the present approach to precision
measurements with atomic ensembles.

C.S. thanks N. Brunner for very helpful comments. This work
was supported by an AI-TF New Faculty Award, an NSERC
Discovery Grant, and the European projects Q-ESSENCE and
EMALI.

%%%%%%%%%%%%%%%%%%%%%%%%%%%%%%%%
%%%%%%%%%%%%%%%%%%%%%%%%%%%%%%%%


\begin{thebibliography}{100}
\bibitem{AAV} Y. Aharonov, D.Z. Albert, and L. Vaidman,
Phys. Rev. Lett. {\bf 60}, 1351 (1988).
\bibitem{foundations} Y. Aharonov, A. Botero, S. Popescu,
B. Reznik, and J. Tollaksen, Phys. Lett. A {\bf 301}, 130
(2002); J.S. Lundeen and A.M. Steinberg, Phys. Rev. Lett.
{\bf 102}, 020404 (2009); K. Yokota, T. Yamamoto, M.
Koashi, and N. Imoto, New J. Phys. {\bf 11}, 033011 (2009).
\bibitem{superluminal} D.R. Solli, C.F. McCormick, R.Y.
Chiao, S. Popescu, and J.M. Hickmann, Phys. Rev. Lett. {\bf
92}, 043601 (2004); N. Brunner, V. Scarani, M.
Wegm\"{u}ller, M. Legr\'{e}, and N. Gisin, Phys. Rev. Lett.
{\bf 93}, 203902 (2004).
\bibitem{networks} N. Brunner, A. Acin, D. Collins, N.
Gisin, and V. Scarani, Phys. Rev. Lett. {\bf 91}, 180402
(2003).
\bibitem{wiseman} H. Wiseman, Phys. Rev. A {\bf 65}, 032111
(2002).
\bibitem{precision} O. Hosten and P. Kwiat, Science {\bf
319}, 787 (2008); P.B. Dixon, D.J. Starling, A.N. Jordan,
and J.C. Howell, Phys. Rev. Lett. {\bf 102}, 173601 (2009);
D.J. Starling, P.B. Dixon, A.N. Jordan, and J.C. Howell,
Phys. Rev. A {\bf 80}, 041803 (2009).
\bibitem{brunner} N. Brunner and C. Simon, Phys. Rev. Lett.
{\bf 105}, 010405 (2010).
\bibitem{starling} D.J. Starling, P.B. Dixon, N.S.
Williams, A.N. Jordan, and J.C. Howell, arXiv:0912.2357.
\bibitem{morelight} N.W.M. Ritchie, J.G. Story, and R.G.
Hulet, Phys. Rev. Lett. {\bf 66}, 1107 (1991); G.J. Pryde,
J.L. O'Brien, A.G. White, T.C. Ralph, and H.M. Wiseman,
Phys. Rev. Lett. {\bf 94}, 220405 (2005); R. Mir {\it et
al.}, New J. Phys. {\bf 9}, 287 (2007); M. Goggin {\it et
al.}, arXiv:0907.1679.
\bibitem{qdots} N.S. Williams and A.N. Jordan, Phys. Rev.
Lett. {\bf 100}, 026804 (2008); A. Romito, Y. Gefen, and
Y.M. Blanter, Phys. Rev. Lett. {\bf 100}, 056801 (2008).
\bibitem{hammerer} K. Hammerer, A.S. S{\o}rensen, and E.S.
Polzik, Rev. Mod. Phys. {\bf 82}, 1041 (2010).
\bibitem{memory} B. Julsgaard, J. Sherson, J.I. Cirac, J.
Fiurasek, and E.S. Polzik, Nature {\bf 432}, 482 (2004).
\bibitem{teleport} J. Sherson {\it et al.}, Nature {\bf
443}, 557 (2006).
\bibitem{briegel} H.-J. Briegel, W. D\"{u}r, J.I. Cirac,
and P. Zoller, Phys. Rev. Lett. {\bf 81}, 5932 (1998).
\bibitem{dlcz} L.-M. Duan, M.D. Lukin, J.I. Cirac, and P.
Zoller, Nature {\bf 414}, 413 (2001).
\bibitem{dlcz-exp} A. Kuzmich {\it et al.}, Nature {\bf
423}, 731 (2003); C.H. van der Wal {\it et al.}, Science
{\bf 301}, 196 (2003); C.W. Chou {\it et al.}, Nature {\bf
438}, 828 (2005); C.W. Chou {\it et al.}, Science {\bf
316}, 1316 (2007); Z.-S. Yuan {\it et al.}, Nature {\bf
454}, 1098 (2008).
\bibitem{sangouard} N. Sangouard, C. Simon, H. de
Riedmatten, and N. Gisin, arXiv:0906.2699, to appear in
Rev. Mod. Phys.
\bibitem{appel} J. Appel {\it et al.}, PNAS {\bf
106}, 10960 (2009).

\end{thebibliography}
\end{document}